\documentclass[conference]{IEEEtran}
\IEEEoverridecommandlockouts
\usepackage{cite}
\usepackage{amsmath,amssymb,amsfonts}
\usepackage{algorithmic}
\usepackage{graphicx}
\usepackage{textcomp}
\usepackage{xcolor}

\def\BibTeX{{\rm B\kern-.05em{\sc i\kern-.025em b}\kern-.08em
    T\kern-.1667em\lower.7ex\hbox{E}\kern-.125emX}}
\begin{document} 

\title{Achieving 45\% efficiency of CIGS/CdS Solar Cell by adding GaAs using optimization techniques
}

\author{\IEEEauthorblockN{Satyam Bhatti, Habib Ullah Manzoor, Ahmed Zoha, and Rami Ghannam} 
\IEEEauthorblockA{\textit{James Watt School of Engineering, University of Glasgow, United Kingdom} \\
\textit{Email: \{s.bhatti.2\}@research.gla.ac.uk}\\
\{rami.ghannam\}@glasgow.ac.uk}
}

\maketitle

\begin{abstract}
This paper proposes an efficient three-layered p-GaAs/p-CIGS/n-CdS (PPN), a unique solar cell architecture. Copper indium gallium selenide (CIGS)-based solar cells exhibit substantial performance than the ones utilizing cadmium sulfide (CdS). On the contrary, CIGS-based devices are more efficient, considering their device performance, environmentally benign nature, and reduced cost. Therefore, our paper proposes a numerical analysis of the homojunction PPN-junction GaAs solar cell structure along with n-ZnO front contact that was simulated using the Solar Cells Capacitance Simulator (SCAPS-1D) software. Moreover, we investigated optimization techniques for evaluating the effect of the thickness and the carrier density on the performance of the PPN layer on solar cell architecture. Subsequently, the paper discusses the electronic characteristics of adding GaAs material on the top of the conventional (PN) junction, further leading to improved values of the parameters, such as the power conversion efficiency (PCE), open-circuit voltage (VOC), fill factor (FF) and short-circuit current density (JSC) of the solar cell. The most promising results of our study show that adding the GaAs layer using the optimised values of thickness as 5 ($\mu$m) and carrier density as $1\times10^{20}$ (1/cm) will result in the maximum PCE, VOC, FF, and JSC of 45.7\%, 1.16 V, 89.52\% and 43.88 $(mA/m^{2})$, respectively, for the proposed solar cell architecture.
\end{abstract}


\section{Introduction}

With the rapid increase in electricity demand, solar cells are playing a vital role in producing green, reliable, and efficient energy sources to meet the United Nation's sustainable development goals-7 \cite{alqallaf2022visualising}. One of the promising solar cell architectures involves the thin film CIGS (Copper Indium Gallium Selenide) and CdS (Copper Sulfide) solar cell, which is well known for yielding higher Power Conversion Efficiencies (PCE) and generates efficient Levelized Cost of Electricity (LCOE) with manufacturing costs minimal as compared to other solar cells architectures \cite{bhatti2022machine, benmir2013analytical}. This paper proposes simulation-based modelling of thin film layer CIGS/CdS solar cells followed by the optimization of the nanowire GaAs (Gallium Arsenide) layer, which was added to the top layer of the baseline multi-junction solar cell architecture. Moreover, the simulations presented in this paper were performed using the Solar Cells Capacitance Simulator (SCAPS-1D), a one-dimensional solar simulation tool that provides an in-depth investigation of electronics and information systems \cite{al2021numerical, mostefaoui2015simulation, bhatti2023machine}. Moreover, SCAPS-1D evaluates one of the vital electrical characteristics such as $PCE$, Fill factor $(FF)$, Current Density curves $(J_{SC})$ and the open circuit voltage $(V_{OC})$ of the solar cells \cite{tan2022numerical}. 

In the recent past, an exponential interest in CIGS and CdS solar cells has recently increased owing to their attractive properties and applications \cite{rawat2023insight}. For the CIGS semiconductor material, the wide direct bandgap range lies between 1.0 to 1.7 eV, whereas that of the CdS lies within the 2.2 and 2.4 eV range and is an important parameter in terms of determining the range wavelengths of light that could be possibly absorbed by the solar cell and help in predicting the PCE of converting the sunlight into electricity \cite{gloeckler2005efficiency}. These factors, specifically the CIGS and CdS semiconductor materials highly favourable materials for photovoltaic energy conversion applications \cite{li2022bto, bhatti2023revolutionizing}. Furthermore, CIGS and CdS materials contribute towards high efficiency, low cost of manufacture, thin-film fabrication technique, environmentally friendly, varied spectral response, high radiation tolerance and absorption coefficient \cite{gulkowski2017experimental}. Therefore, research in these materials has led to the advanced development and fabrication of semiconductor devices, optoelectronics, and nanotechnology from photovoltaic to biosensors and are widely used as catalysts in chemical reactions involving hydrogen production from water \cite{9970858}. 

On the contrary, one limiting factor in developing more efficient solar cells using semiconductor materials lies in their incapability to absorb light from the solar spectrum \cite{ramanujam2017copper}. Following this, CdS also led to a concern about toxicity and stability because of the availability of Cadium in the solar cells \cite{kosyachenko2013optical}. However, mechanical stacking of multijunction of semiconductor materials can help to mitigate these concerns whilst enhancing the overall solar cell efficiency \cite{ramanujam2020flexible}. In our paper, we run a number of simulations involving the traditional single-junction CIGS/CdS solar cell layers and then later add the GaAs layer to study the properties of the multijunction solar cell. From the literature, several attempts have been made to optimize the overall efficiency of the solar cell and were conducted by \cite{persson2006thin, nishinaga2020crystalline, fatemi2020analysis}. The main idea behind this analysis is the improvement of the device efficiency using materials cheaper than conventional CIGS \cite{khan2018optimization}. A 5 $\mu$m of a new layer p-GaAs has been added for that purpose. Various thicknesses of the CIGS absorber layer ranging from 0.5 to 5 $\mu$m have been used. 

Accordingly, the findings of our study showcase that an increase in the absorber layer thickness improves the performance and overall power conversion efficiency of the new CIGS solar cells. Initially, the study incorporated a window layer of ZnO, a buffer layer (CdS), an absorbent layer (CIGS) and a GaAs layer with varied values of thicknesses between, 0.5, 1, 1.5, 2, 2.5, 3, 3.5, 4, 4.5 and 5 $\mu$m. Also, the optimisation of the GaAs material is performed with the help of the heatmap confusion matrix in order to analyse the most optimized thickness and carrier density of different layers of solar cells. In addition, a comparison of results including the performance and efficiency of the PN-junction (optimized CIGS and CdS) solar cell was estimated. The most promising results of the paper revealed that a thin top layer of p-GaAs on the conventional solar cell (p-CIGS and n-CdS) with an optimized thickness layer and high carrier density had a considerable influence on the performance of the solar cell architecture. Adding a p-GaAs layer as thin as 5 $\mu$m on the top of the PN-junction solar cell substantially improved the conversion efficiency of the solar cell from 20.07\% (unoptimized PN) to 45.47 \% (optimized PPN). The results showed that the new ultra-thin CIGS solar cells structure has performance parameters that are comparable to those of the conventional ones with reduced cost.

Therefore, our paper proposes a simple three-layered p-GaAs/p-CIGS/n-CdS (PPN) solar cell with a practical thickness and a high conversion efficiency. The PPN solar cell comprises a high p-GaAs composition on top of p-CIGS and n-CdS multi-junction solar cells. In addition, computing a thin p-GaAs layer created a graded energy bandgap and the solar PN junction could only absorb photons having an energy equal to or greater than the energy bandgap. Herein, it is worth mentioning that a graded energy bandgap has a wider energy bandgap to absorb more photons. Moreover, adding an extra thin layer increased the number of holes in the solar cells, ultimately increasing power conversion efficiency. The behaviour and performance of the solar cells were evaluated by performing a comprehensive set of simulations under different configurations (i.e., thickness and carrier density of the layer) using the SCAPS-1D software.

\subsection{Organisation of the Paper}

Our paper is divided into 8 Sections. Section 2 of the paper discusses the significance, need and limitations of the CIGS/CdS multijunction solar cell architecture. Followed by, Section 3 describes the process of setting up our simulation environment in the SCAPS-1D. Section 4 includes the efficiency optimization of the CIGS/CdS solar cell with the help of thickness and carrier density optimization using the heatmap confusion matrix. Section 5 presents the results of electrical parameters after adding a GaAs layer on top of the CIGS/CdS solar cell. Next, Section 6 discusses a critical comparison of IV, PV, and QE characteristics and the high-temperature effect. Lastly, section 7 discusses the obtained results in detail, and concluding remarks are presented in Section 8.

\begin{figure}[!t]
 \centering
 \includegraphics[width=8cm]{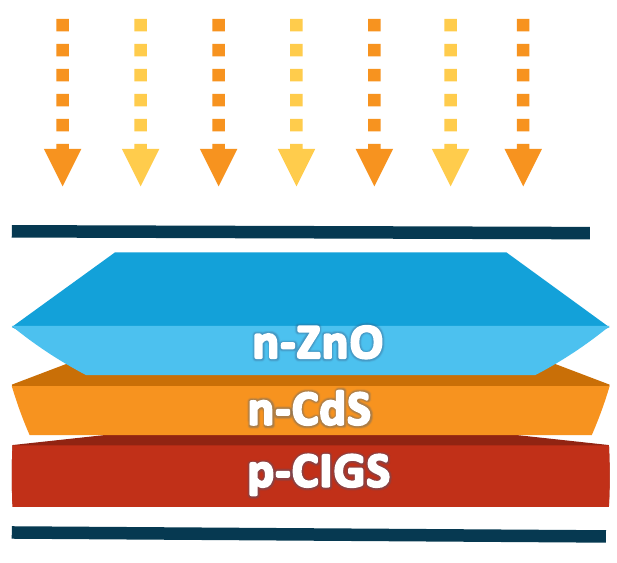}
 \caption{Figure represents the baseline structure of the proposed multi-junction solar cell architecture, with simple three p-CIGS/n-CdS/n-ZnO, where the ZnO layer acts as a window/buffer layer. The solar light in the structure is incident from the right contact (front) using the simulation tool SCAPS-1D. The baseline values for the given solar cell architecture are set as mentioned in the simulation setup whilst the thickness and current density (doping) values for the baseline are set as 0.5 $\mu$m and $1\times10^{10}$ (1/$cm^3$), respectively. }
 \label{fig:Ref1}
\end{figure}

\subsection{Contributions to the Research Paper}

The key contributions to the research article are presented as follows:

\begin{table*}[ht]
\caption{The table below depicts the input parameters that were set for the simulation setup.}
\begin{center}
\begin{tabular}{|c|c|c|c|c|c|c|}
\hline
\multicolumn{1}{|c|}{\raisebox{-1.50ex}[0cm][0cm]{\!\textbf{Input Electrical Parameters}\!}}
& \multicolumn{1}{|c|}{\textbf{{Measurement}}}
& \multicolumn{1}{|c|}{\textbf{{Layer 1:}}}
& \multicolumn{1}{|c|}{\textbf{{Layer 2:}}}
& \multicolumn{1}{|c|}{\textbf{{Layer 3:}}}
& \multicolumn{1}{|c|}{\textbf{{Layer 4:}}} \\
& \textbf{Unit} & \textbf{p-GaAs} & \textbf{p-CIGS} & \textbf{n-CdS} & \textbf{n-ZnO} \\ \hline
\textbf{Bandgap} & (eV)                              & $1.420$       &  $1.100$        &  $2.450$   &  $3.300$  \\ \hline
\textbf{Electron Affinity} & (eV)                    & $4.070$       &  $4.500$        &  $4.400$   &  $4.600$  \\ \hline
\textbf{Dielectric Permittivity}  &  (relative)        & $12.900$      &  $13.600$       &  $10.000$   &  $9.000$  \\ \hline
\textbf{Conduction Band Effective Density of states} & (1/$cm^3$) & $2\times10^{18}$   &  $2.2\times10^{18}$  &  $2.2\times10^{18}$   &  $2.2\times10^{18}$  \\ \hline
\textbf{Valence Band Effective Density of states} & (1/$cm^3$) & $1\times10^{19}$   &  $1.8\times10^{19}$  &  $1.8\times10^{19}$   &  $1.8\times10^{19}$  \\ \hline
\textbf{Electron Thermal Velocity}  & (cm/s)          & $1\times10^{7}$	  &  $1\times10^{7}$	    &  $1\times10^{7}$	 &  $1\times10^{7}$ \\ \hline
\textbf{Hole Thermal Velocity} & (cm/s)              & $1\times10^{7}$	  &  $1\times10^{7}$	    &  $1\times10^{7}$	 &  $1\times10^{7}$ \\ \hline
\textbf{Electron Mobility}  & ($cm^2$/Vs)             & $1\times10^{3}$	  &  $1\times10^{2}$		&  $1\times10^{2}$	 &  $1\times10^{2}$ \\ \hline   
\textbf{Hole Mobility} &   ($cm^2$/Vs)                 & $1\times10^{2}$	  &  $1\times10^{1}$		&  $1\times10^{1}$	 &  $2.5\times10^{1}$ \\ \hline   
\end{tabular}
\label{table1}
\end{center}
\end{table*}

\begin{itemize}

    \item Discussed the significance, need and limitations along with a thorough literature review of the CIGS and CdS multijunction solar cell. 

     \item Implemented simulations on the SCAPS-1D tool for designing the most optimized, high-efficiency and robust solar cell architecture.

     \item Performed efficiency optimization for the CIGS/CdS multi-junction solar cell architecture by adjusting the thickness and carrier density using the heatmap confusion matrix.

    \item Whilst performing simulations, critical optimization techniques are applied to the CIGS/CdS solar cell as a first step and then, for CIGS/CdS/GaAs as a second step with the help of the heatmap confusion matrix ranging from 0.5 $\mu$m to 5 $\mu$m for thickness and 10 (1.00En) (1/$cm^3$) and 20 current density, respectively.
     
     \item In the second step, introduced a novel multi-junction solar cell architecture by adding an n-GaAs layer on top of the pre-existing CIGS/CdS multi-junction solar cell and carried out the thickness and carrier density optimization to achieve the highest PCE value. 

    \item Further, investigated the electronic and electrical characteristics such as PCE, fill factor, current density and open circuit voltage of the p-CIGS/p-CdS/n-GaAs multijunction solar architecture. 

    \item Lastly, a thorough comparative analysis is presented, showing the IV, PV and QE characteristics graphs for the most optimized solar cell architecture with their respective optimized thickness and current density values.

\end{itemize}

\section{CIGS and CdS Multijunction Solar Cell}

Initially, a multijunction solar cell architecture using the CIGS and CdS semiconductor materials is designed. CIGS and CdS materials are widely used for manufacturing purposes, mainly because of key parameters such as low production cost and high solar energy yield. The literature suggests that the materials have achieved higher PCE values than their counterpart Silicon solar cells installed on household rooftops. Moreover, the CIGS solar cells use a thin layer of CIGS as an absorbing layer and have high absorption coefficient values, which eventually results in high-efficiency solar cells. The flexible nature of the CIGS semiconductor materials broadens their application in various industries, research and industry perspectives such as photovoltaic, biosensors, portable electronic devices, off-grid power systems, utility-scale solar power plants, for powering remote sensing and for communication devices \cite{bhatti2021estimation}.

Subsequently, the semiconductor CdS is used as the window or also known as the buffer layer in the solar cells. The window layer is defined as a thin, transparent allowing the sunlight to pass through the absorber layer and also restricts the unwanted recombination of the holes and electrons generated by the p-CIGS (absorbed photons) layer. Moreover, the window layer is typically made of a transparent conducting oxide (TCO), such as indium tin oxide (ITO), fluorine-doped tin oxide (FTO), or zinc oxide (ZnO). These materials have high transparency to visible light and low electrical resistance, which allows them to efficiently collect and transport the electrons generated by the absorbed photons. Furthermore, the choice of window layer material depends on several factors, including the specific absorber material used in the solar cell, the desired efficiency, and the cost of the materials. Different TCOs have different properties that can affect the performance of the solar cell, such as their work function, surface roughness, and doping level.

Therefore, the window layer plays a critical role in the performance of a solar cell by controlling the flow of charge carriers and minimizing energy losses due to recombination, while also allowing sunlight to pass through to the absorber layer. Accordingly, in the multijunction solar cell of p-CIGS/p-CdS, CdS is often used as the window layer in the respective solar cell architecture. Additionally, the transparent window layer sits on top of the absorbing layer and allows light to pass through to reach the CIGS layer. CdS has a high optical transmission, which allows it to efficiently transmit light to the CIGS layer, while also helping to protect the CIGS layer from degradation due to exposure to air and moisture. However, it's important to note that cadmium is a toxic substance, raising concerns about CdS solar cells' environmental impact. Subsequently, efforts are being made to develop alternative materials for window layers, such as zinc oxide (ZnO), a non-toxic alternative to CdS.

\subsection{CIGS/CdS: Significance and Need}

Previous research studies indicate that the combination of CIGS and CdS multijunction solar cells is known to provide promising electrical and electronic characteristics even in harsh environments along with achieving considerably higher efficiencies. CIGS and CdS solar cell layers are used due to the fact that they offer numerous advantages over the widely existing materials manufactured in the solar industry. Not only, CIGS has high efficiency in converting sunlight into electricity but also it has the capacity to generate more power (Ws) per unit of surface area (A) in comparison with other materials. Accordingly, this makes CIGS and CdS materials ideal for applications in solar cells where either the space is limited or reduced size is targeted, for example, rooftops or portable devices. In addition, CIGS materials are known to be flexible, which allows them to be used in various applications, including curved or irregular surfaces. Moreover, this property also makes the design of solar cells suitable for building-integrated photovoltaics (BIPV) where they are integrated into the design of buildings and other complex structures. Lastly, CdS materials proved to be an excellent window or buffer layer material due to their high optical transmission, which allows them to efficiently transmit light to the CIGS layer while also providing protection from environmental factors that can degrade the solar cell \cite{lee2015effect}. Overall, the combination of CIGS and CdS offers a high-efficiency, flexible, and durable solution for solar cells that can be used in a wide range of applications.

\begin{figure*}[!t]
 \centering
 \includegraphics[width=18cm]{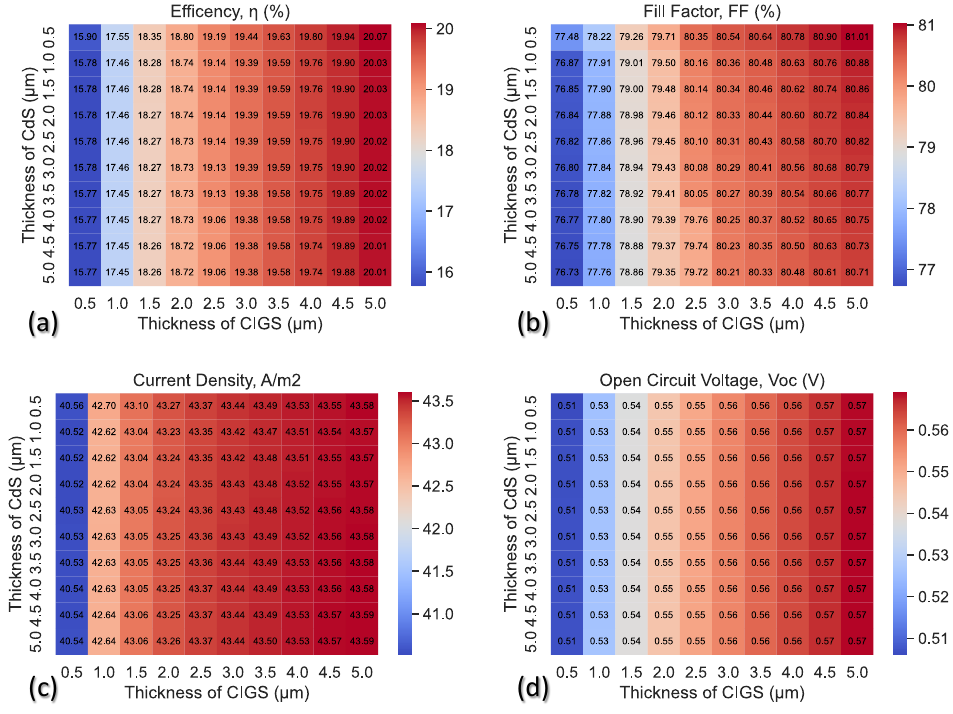}
 \caption{Figure represents the optimization technique used to evaluate the most efficient value of the thickness for the p-CIGS and n-CdS semiconductor materials using the heatmap confusion matrix. In addition, (a) Outputs the Efficiency, $\eta$ (\%), (b) Outputs the Fill Factor, FF (\%), (c) Current Density (A/$cm^3$), (d) Open Circuit Voltage ($V_{OC}$), (e) Comparison of the IV and PV characteristics after the thickness optimization, (f) Quantum Efficiency (\%) Curve.}
 \label{fig:Ref2}
\end{figure*}

\subsection{CIGS/CdS: Limitations}

While CIGS and CdS solar cells offer many advantages, there are also some challenges associated with their use. One challenge is the cost of producing CIGS solar cells. The materials used to make CIGS solar cells are relatively expensive, making the manufacturing process more costly than other types of solar cells. This has limited the widespread adoption of CIGS solar cells, especially in utility-scale applications. Another challenge is the potential environmental impact of using CdS as a window layer material. Cadmium is a toxic substance, and concerns have been raised about the possibility of environmental contamination from CdS solar cells during their production, use, and disposal. Efforts are being made to develop non-toxic alternatives to CdS for use in window layers, such as zinc oxide (ZnO), but these materials are not yet widely used in commercial applications. Finally, CIGS solar cells are susceptible to degradation over time due to exposure to moisture and air. This can reduce the efficiency of the solar cell and limit its lifespan. However, research is ongoing to develop new encapsulation methods and materials that can protect the CIGS layer from degradation and extend the lifespan of CIGS solar cells. Overall, while there are some challenges associated with using CIGS and CdS solar cells, ongoing research and development efforts are aimed at addressing these challenges and improving the efficiency and durability of these solar cell technologies \cite{9971065}.


\begin{figure*}[ht]
 \centering
 \includegraphics[width=18cm]{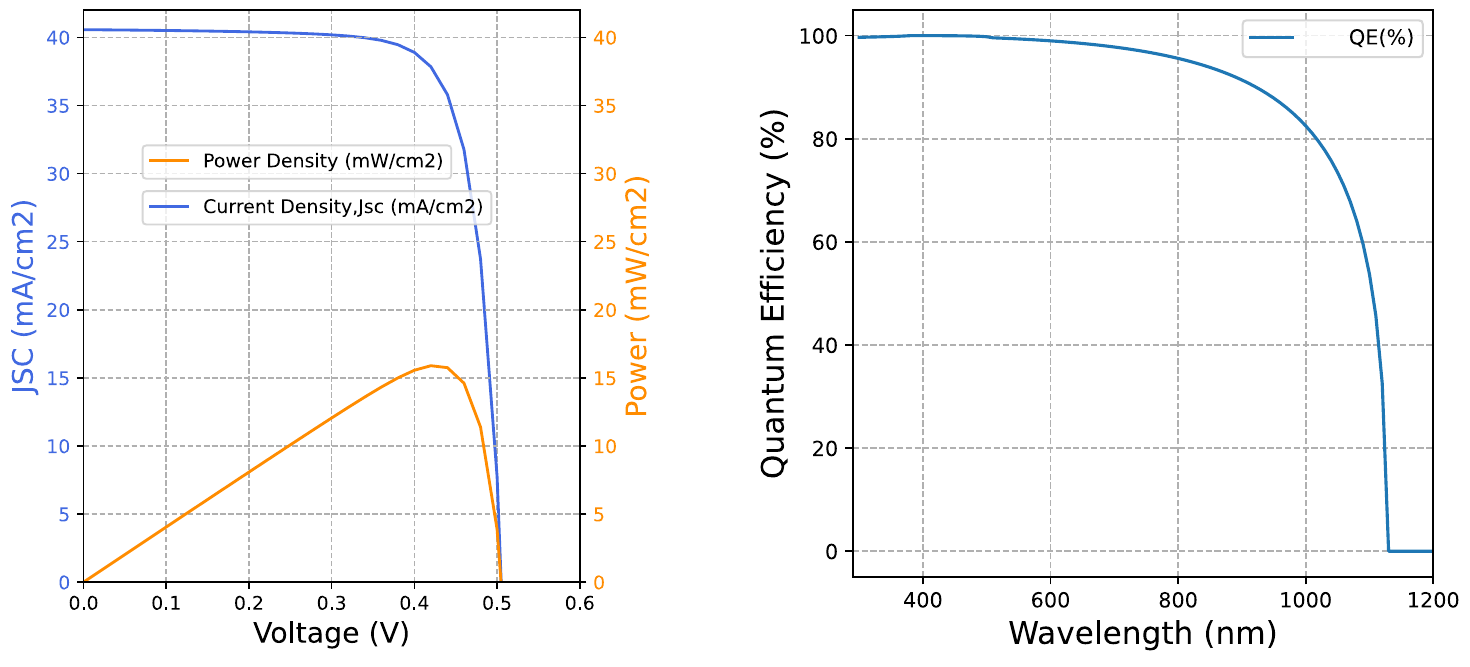}
 \caption{Figure represents the optimization technique used to evaluate the most efficient value of the thickness for the p-CIGS and n-CdS semiconductor materials using the heatmap confusion matrix. In addition, (a) Outputs the Efficiency, $\eta$ (\%), (b) Outputs the Fill Factor, FF (\%), (c) Current Density (A/$cm^3$), (d) Open Circuit Voltage ($V_{OC}$), (e) Comparison of the IV and PV characteristics after the thickness optimization, (f) Quantum Efficiency (\%) Curve.}
 \label{fig:Ref4}
\end{figure*}

The overall efficiency of CIGS and CdS solar cells has improved significantly over the past few decades, making them competitive with other types of solar cells in terms of efficiency. The current record for CIGS solar cell efficiency is around 23.35\, which is close to the efficiency of traditional silicon-based solar cells. The high efficiency of CIGS solar cells is due to their ability to absorb a broad range of wavelengths of light, including the blue and green parts of the spectrum, which traditional silicon solar cells cannot absorb efficiently. The CdS window layer in CIGS solar cells also plays a critical role in their overall efficiency. CdS has high optical transmission, which allows it to efficiently transmit light to the CIGS layer, resulting in higher conversion efficiencies. However, as mentioned earlier, the use of CdS raises concerns about its potential environmental impact. Overall, the efficiency of CIGS and CdS solar cells is highly competitive with other types of solar cells, and ongoing research and development efforts are focused on improving their efficiency and reducing their cost.

The efficiency of CIGS and CdS solar cells can be evaluated in both simulation-based environments and real-world environments. In simulation-based environments, the efficiency of solar cells can be modeled using computer simulations that take into account various factors, such as material properties, cell design, and environmental conditions. These simulations can provide insights into the theoretical efficiency of solar cells and can help guide the design and optimization of solar cell materials and structures. In real-world environments, the efficiency of solar cells can be evaluated by measuring their performance under actual operating conditions. Factors such as temperature, humidity, and shading can affect the performance of solar cells in real-world environments, leading to variations in efficiency compared to simulation-based results. Generally, the efficiency of CIGS and CdS solar cells in real-world environments is lower than in simulation-based environments due to various factors such as partial shading, thermal losses, and other environmental factors. However, ongoing research and development efforts are focused on improving the real-world performance of CIGS and CdS solar cells, including developing new encapsulation techniques, optimizing the cell design, and developing new materials for use in the window layer. Overall, the efficiency of CIGS and CdS solar cells is evaluated in both simulation-based and real-world environments, and ongoing research is focused on improving their efficiency and performance in both environments.


\begin{figure*}[!t]
 \centering
 \includegraphics[width=18cm]{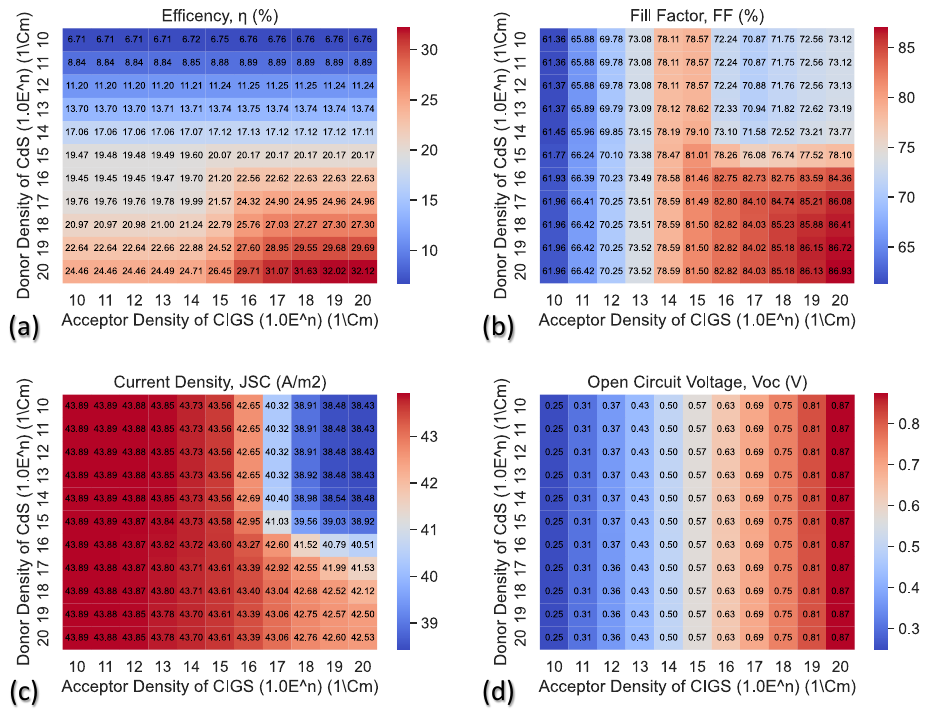}
 \caption{Figure represents an optimization of the carrier charge density, also known as doping concentration (1/$cm$) for the p-CIGS/n-CdS/n-ZnO multi-junction solar cell architecture incorporating the Heatmap confusion matrix. Likewise to the thickness optimization, Doping optimisation values are as  follows: (a) Outputs the Efficiency, $\eta$ (\%), (b) Outputs the Fill Factor, FF (\%), (c) Current Density (A/$cm^3$), (d) Open Circuit Voltage ($V_{OC}$), (e) Comparison of the IV and PV characteristics after the thickness optimization, (f) Quantum Efficiency (\%) Curve.}
 \label{fig:Ref6}
\end{figure*}

\section{Simulation Setup}

For the analysis of the different solar cell architectures, a simulation environment was developed on the SCAPS-1D tool for examining the electronic and electrical parameters along with measuring the PCE, FF, VOC and JSC values. Initially, a solar cell was designed using the three-layered P-N-N semiconductor layers. With light incident from the right contact (front), followed by a layer of n-ZnO, n-CdS and p-GaAs to the left contact (back) formed a multi-junction solar cell architecture. The incident light plays a pivotal role in determining the efficiency of the solar cell and thus, in our simulation setup, we introduced the incident light from the right contact (front) throughout the simulations settings. Furthermore, the initial working point of the simulations was set as follows: Temperature - 300K, Voltage - 0V, Frequency - $10^6$ Hz and the number of points as 5.

In addition to this, for plotting the electrical characteristics curves, the settings for Pause at each stop were applied as V1 (0V to -0.8V), V2 (0.8V to 0.8V), frequency (f1: $10^2$ Hz to f2: $10^6$ Hz) and the wavelength (WL1: 300nm to WL2: 900nm). Moreover, at each step, the number of points was set as 41, 81, 21, and 61 with an increment of 0.02V, 0.02V, 5 points per decade, and 10nm, respectively. It is worth mentioning that the author later in the study also introduced a layer of GaAs with the aim of achieving the highest efficiency and accordingly, all the settings of the working point remain the same for different multi-junction solar cell architecture. Subsequently, figure 1 represents the direction of the incident light along with the p-CIGS/n-CdS/n-Zno multi-junction solar cell architecture and Table 1 depicts the input electrical parameters such as the bandgap, electron affinity (eV), dielectric permittivity (relative), conduction band effective density of states, valence band effective density of states, electron thermal velocity, hole thermal velocity, electron and hole mobility of the respective, 4 layered solar cell architecture.

\begin{figure*}[!t]
 \centering
 \includegraphics[width=18cm]{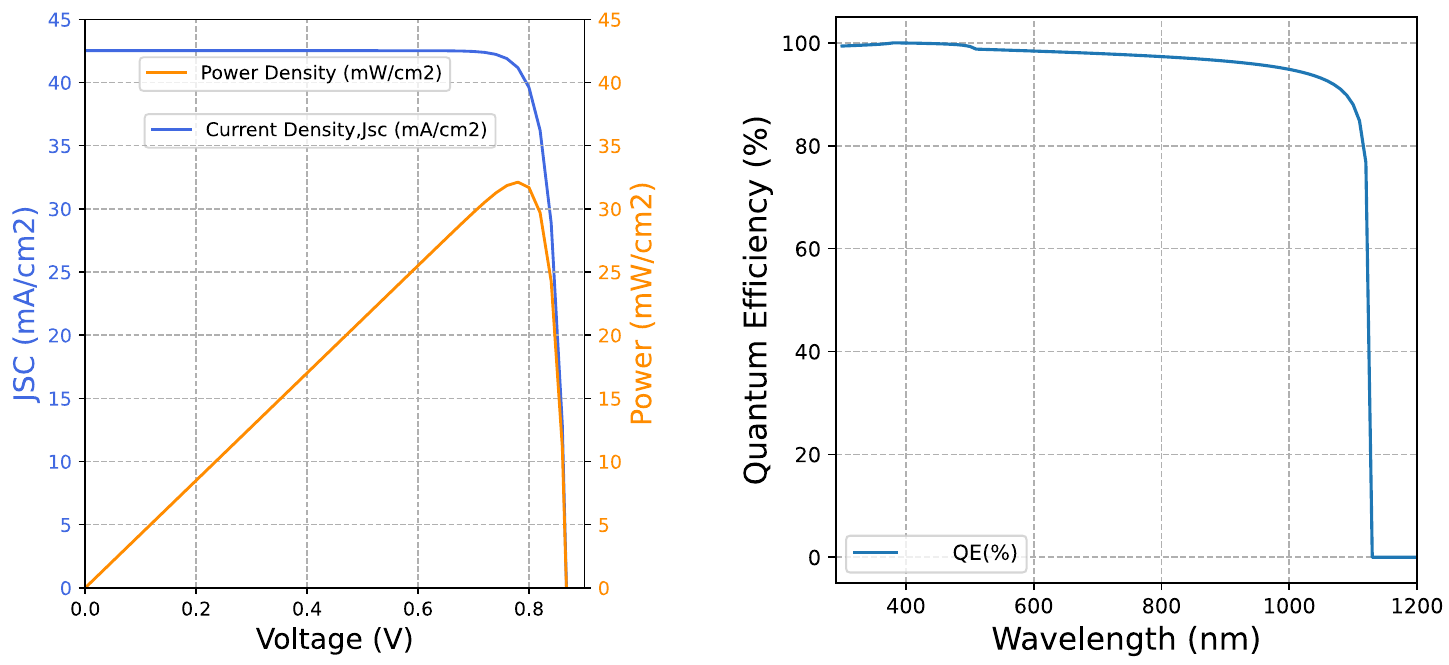}
 \caption{Figure represents an optimization of the carrier charge density, also known as doping concentration (1/$cm$) for the p-CIGS/n-CdS/n-ZnO multi-junction solar cell architecture incorporating the Heatmap confusion matrix. Likewise to the thickness optimization, Doping optimisation values are as  follows: (a) Outputs the Efficiency, $\eta$ (\%), (b) Outputs the Fill Factor, FF (\%), (c) Current Density (A/$cm^3$), (d) Open Circuit Voltage ($V_{OC}$), (e) Comparison of the IV and PV characteristics after the thickness optimization, (f) Quantum Efficiency (\%) Curve.}
 \label{fig:Ref7}
\end{figure*}

\section{Efficiency Optimization}

After incorporating the necessary input electrical values and setting up the simulation environment, in the next step, the authors conducted the efficiency optimization consisting of three important steps of the study. Accordingly, to meet the objective of this paper, a critical optimization of different solar cell architectures is performed. The first step of optimization involves varying the thickness of the CIGS and CdS layers in the multi-junction solar cell of p-CIGS/n-CdS/n-ZnO. Followed by, the second step included the same solar cell architecture, however, the efficiency optimization is evaluated with the help of changing the doping values. Lastly, the third step introduces a GaAs layer to the existing solar cell architecture and accordingly, both the thickness and doping optimization is performed simultaneously.

\subsection{STEP 1: Optimization of Thickness for CIGS and CdS layer}

Initially, a baseline solar cell architecture is proposed consisting of p-CIGS/n-CdS/n-ZnO and as a first step, the thickness optimization is performed for estimating the overall efficiency of the solar cell at different values of thickness that range from 0.5 to 5.0 $\mu$m, at an increment of 0.5 $\mu$m. Accordingly, the increment in the values of thickness is performed in such a manner that an increment is made for each 0.5 $\mu$m value for both the CIGS and CdS semiconductor material of the solar cell independently. Therefore, to present a relation between the two parameters' increment of thickness, the authors used the heatmap correlation matrix to provide a clear indication of the values of the overall performance of the solar cell at varied values of thickness of both materials.

The estimation of thickness is highly essential for designing the most optimized solar cell due to cost saving of materials and subsequently, the solar cell manufacturers could use the materials optimally. The results from the heatmap confusion matrix showcase four parameters as Efficiency, fill factor, current density and open circuit voltage. For thickness optimization of the materials p-CIGS/n-CdS, the heatmap confusion matrix indicates a thickness value of 0.5 $\mu$m for CdS and 5.0 $\mu$m for CIGS, giving a maximum efficiency of solar cell design as 20.07\%. Followed by a thickness of 1 $\mu$m for CdS and 5.0 $\mu$m for CIGS yields a maximum Fill factor of 80.88\%. Whereas the maximum values of parameters current density and open circuit voltage output a 43.59 $A/m^2$ and 0.57 V, respectively for thickness values of 4.5 $\mu$m for CdS and 5.0 $\mu$m for CIGS.

In addition, for the electrical performance of the proposed solar cell, PV, IV and QE characteristics are plotted to measure the maximum power point tracking of the solar cell. Accordingly, the maximum value as observed from the IV curve is 45 $A/m^2$ and the maximum power as analysed from the PV curve is 32 W. Subsequently, the maximum quantum efficiency of the baseline proposed solar cell architecture is 95\%.

\subsection{STEP 2: Optimization of Carrier Density for CIGS and CdS layer}

After optimizing the thickness values of the p-CIGS/n-CdS/n-ZnO multi-junction solar cell architecture, the authors conducted the optimization of the carrier density parameter of the solar cell using the same input values as set in the working simulation environment. However, the authors used the most optimized value of the thickness of the solar cell as calculated in the above subsection, i.e., 0.5 $\mu$m for CdS semiconductor material and 5.0 $\mu$m for CIGS material. It is worth mentioning that the authors used this value of thickness for the remaining simulations of the study.

\begin{figure}[!t]
 \centering
 \includegraphics[width=8cm]{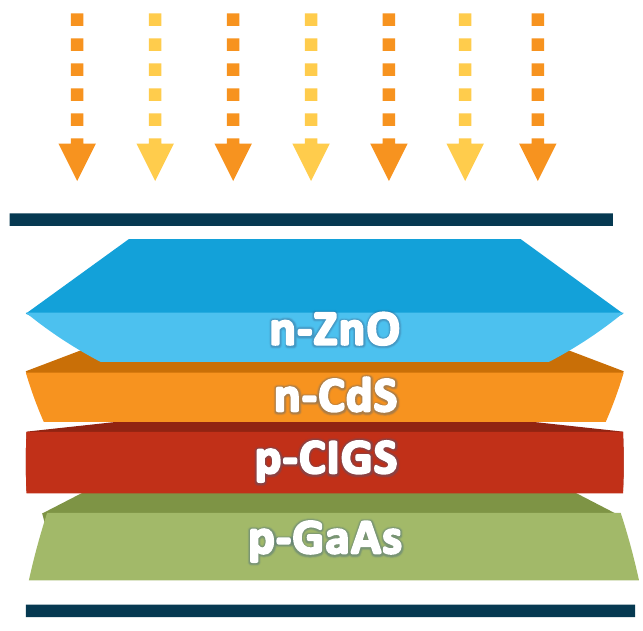}
 \caption{Figure represents the introduction of the p-GaAs layer to the proposed baseline multi-junction solar cell architecture. The GaAs layer is added on top of the p-CIGS layer next to the left contact (back) of the proposed solar cell architecture for analysing the electrical and electronic performance in the real-world environment with the help of the SCAPS-1D simulation tool. Moreover, the light in this scenario is incident from the right contact (front) to achieve the maximum possible efficiency of the proposed solar cell architecture.}
 \label{fig:Ref8}
\end{figure}

Furthermore, the optimization of the acceptor density of the CIGS $1.0\times10^n$ ($A/m^2$) and the donor density of the CdS $1.0\times10^n$ ($A/m^2$) is evaluated using the heatmap confusion matrix which gives the values of the doping concentration for both the CdS and CIGS. Likewise, to the previous subsection, herein, also the authors measured the electrical characteristics such as Efficiency (\%), Fill Factor (\%), Current Density ($A/m^2$) and the open circuit voltage (V) for the critical optimization of the doping concentration. Accordingly, figure 4 (a), (b) and (d) indicate the optimization of the doping concentration at $1\times10^{20}$ (1/cm) for both CIGS and CdS materials yields a maximum value of efficiency, fill factor and open circuit voltage as 32.12 \%, 86.93 \% and 0.87 V, respectively. On the contrary, the optimized value of doping concentration for donor density of CdS lies in the range of $1\times10^{10}$ (1/cm) to $1\times10^{20}$ (1/cm) at an acceptor density of CIGS as $1\times10^{10}$ (1/cm) resulted in a maximum value of current density equivalent to 43.89 ($A/m^2$). 

\begin{figure*}[ht]
 \centering
 \includegraphics[width=18cm]{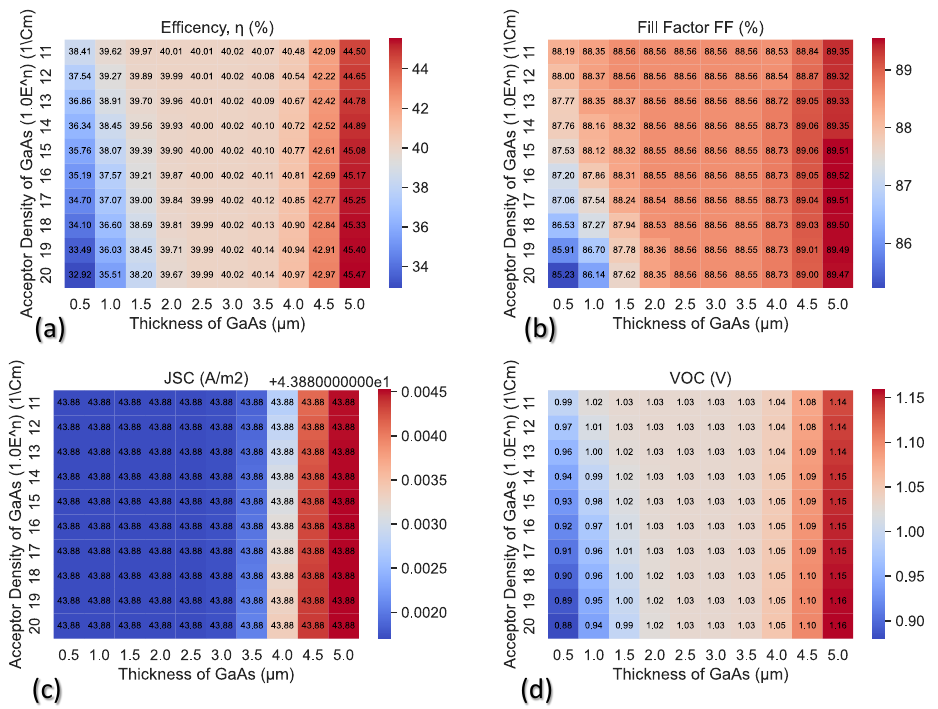}
 \caption{Figure represents the output results for the simulations performed by introducing the GaAs layer to the proposed baseline solar cell architecture. Herein, the optimization for the thickness and the charge current density (doping concentration) is evaluated simultaneously using the same heatmap confusion matrix. Accordingly, electrical characteristics are presented as: (a) Outputs the Efficiency, $\eta$ (\%), (b) Outputs the Fill Factor, FF (\%), (c) Current Density (A/$cm^3$), (d) Open Circuit Voltage ($V_{OC}$), (e) Comparison of the IV and PV characteristics after the thickness optimization, (f) Quantum Efficiency (\%) Curve, that includes the GaAs layer to the proposed multi-junction solar cell. }
 \label{fig:Ref9}
\end{figure*}

In addition to this, the electrical characteristics are calculated from the SCAPS-1D simulation tool. Figure 5 showcases the IV, PV and quantum efficiency for the most optimized values of the doping concentration values at $1\times10^{10}$ (1/cm) for the donor density of CdS and $1\times10^{20}$ (1/cm) for the CIGS semiconductor materials. Accordingly, the curves indicate that the maximum value of current density is 43.2 $A/m^2$ and the maximum power is 32.91 W. Subsequently, the maximum value of the quantum efficiency at the optimized inputs of doping concentration results in 96\%.

\begin{figure*}[!t]
 \centering
 \includegraphics[width=18cm]{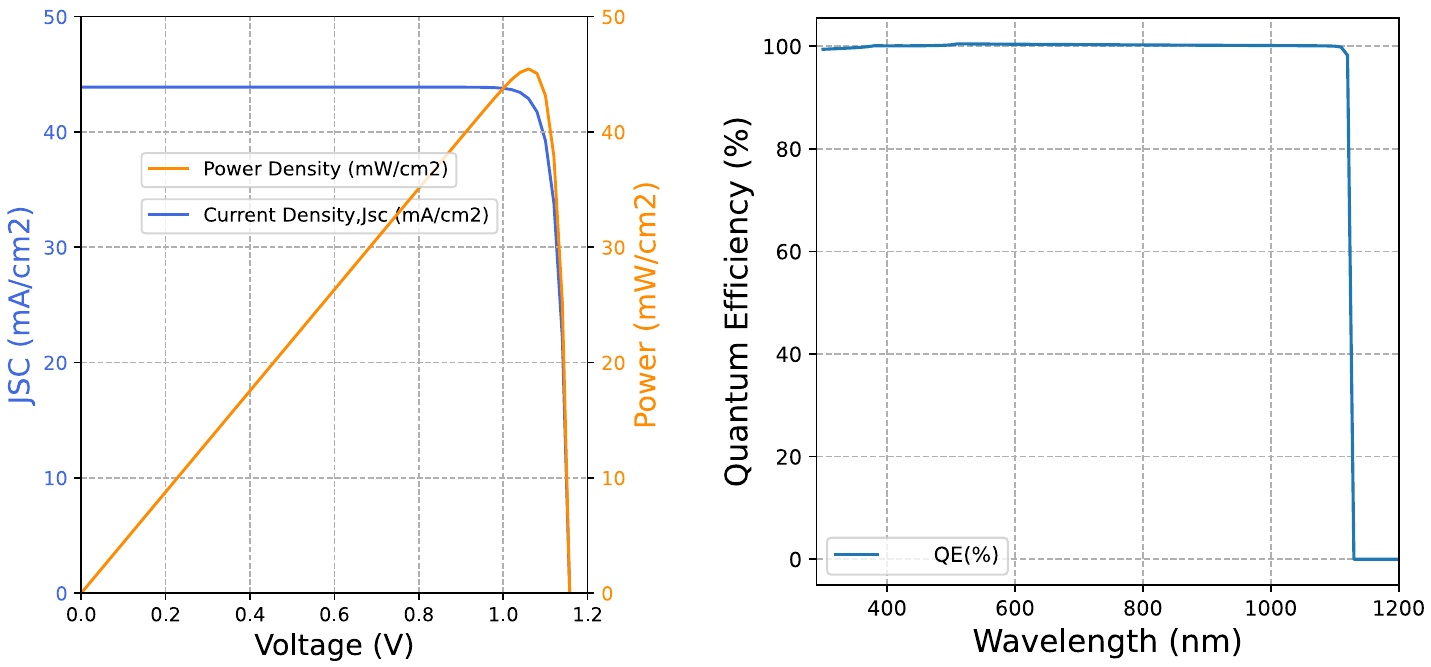}
 \caption{Figure represents the output results for the simulations performed by introducing the GaAs layer to the proposed baseline solar cell architecture. Herein, the optimization for the thickness and the charge current density (doping concentration) is evaluated simultaneously using the same heatmap confusion matrix. Accordingly, electrical characteristics are presented as: (a) Outputs the Efficiency, $\eta$ (\%), (b) Outputs the Fill Factor, FF (\%), (c) Current Density (A/$cm^3$), (d) Open Circuit Voltage ($V_{OC}$), (e) Comparison of the IV and PV characteristics after the thickness optimization, (f) Quantum Efficiency (\%) Curve, that includes the GaAs layer to the proposed multi-junction solar cell. }
 \label{fig:Ref10}
\end{figure*}

\section{Introducing GaAs Layer}

One of the most common semiconductor materials used in the multi-junction solar cell is Gallium Arsenide (GaAs) which is primarily stacked on the top of either each layer or at the top of all layers proposed in a solar cell architecture.

Each layer is designed to absorb a specific portion of the solar spectrum, allowing the cell to capture a broader range of sunlight and convert it into electricity more efficiently. One common material used in multi-junction solar cells is Gallium Arsenide (GaAs). GaAs is a semiconductor material with unique properties that make it well-suited for solar cell applications. Here's why GaAs are used as a semiconductor layer in multi-junction solar cells:

\begin{itemize}

    \item \textbf{High energy conversion efficiency:} GaAs have a relatively high energy conversion efficiency compared to other semiconductor materials used in solar cells. It has a direct bandgap, which means it can efficiently convert sunlight into electricity without losing much energy as heat.

    \item \textbf{Wide bandgap:} GaAs has a wide bandgap, which allows them to absorb higher-energy photons from the solar spectrum. By incorporating GaAs in the solar cell stack, it can absorb photons from the blue and green regions of the spectrum, which are not efficiently absorbed by other materials such as silicon (commonly used in single-junction solar cells).

    \item \textbf{Tandem cell configuration:} In a multi-junction solar cell, the semiconductor layers are arranged in a tandem configuration, with each layer tuned to absorb a specific part of the solar spectrum. GaAs are often used as the top layer in the stack because it has a higher bandgap than other materials, making it suitable for capturing the higher-energy photons. The layers beneath the GaAs layer can be designed to absorb lower-energy photons, ensuring efficient use of the entire solar spectrum.

    \item \textbf{Temperature stability:} GaAs have excellent temperature stability, allowing them to maintain their high performance even at elevated temperatures. This characteristic is crucial for solar cells, as they can heat up under intense sunlight.

    \item \textbf{Mature technology:} GaAs has been extensively researched and developed for various applications, including solar cells. It benefits from a well-established manufacturing process and has a proven track record in high-performance photovoltaic devices \cite{bhatti2023ambient}.

    \item \textbf{High electron mobility:} GaAs have a higher electron mobility compared to other common semiconductor materials like silicon. This property makes GaAs suitable for high-speed electronic devices, such as field-effect transistors (FETs) and integrated circuits, where fast switching and high-frequency operation are required.

    \item \textbf{Low noise characteristics:} GaAs exhibit low noise characteristics, making them ideal for applications in low-noise amplifiers and microwave devices. This property is particularly advantageous in high-frequency communication systems and radar technology.

    \item \textbf{Wide frequency range:} GaAs exhibit excellent performance across a wide frequency range, including microwave and millimetre-wave frequencies. It enables the development of devices and circuits for wireless communications, satellite communications, radar systems, and high-frequency electronics.

    \item \textbf{High power handling capability:} GaAs materials can handle high power levels without significant degradation in performance. This property makes GaAs suitable for power amplifiers and other high-power electronic devices, including those used in telecommunications and defence applications.

    \item \textbf{Optoelectronic applications:} GaAs is widely used in optoelectronic devices such as light-emitting diodes (LEDs), laser diodes, and photodetectors. GaAs-based LEDs and laser diodes have superior performance in terms of efficiency, brightness, and wavelength range, making them valuable for applications in lighting, optical communications, and optical sensing.

    \item \textbf{Compatibility with complementary metal-oxide-semiconductor (CMOS) technology:} GaAs can be integrated with CMOS technology, allowing for the development of hybrid circuits and systems that leverage the advantages of both GaAs and CMOS. This integration enables the fabrication of high-performance, mixed-signal devices and integrated circuits with diverse functionality.

    \item \textbf{Radiation hardness:} GaAs exhibit inherent radiation hardness, meaning they can withstand the effects of ionizing radiation without significant degradation in performance. This characteristic makes GaAs suitable for applications in space technology, nuclear power plants, and high-energy physics experiments.

\end{itemize}

By incorporating a GaAs semiconductor layer in a multi-junction solar cell, the overall efficiency of the cell can be significantly increased. GaAs help capture a broader range of sunlight, including higher-energy photons, and convert them into electricity more effectively, leading to improved solar cell performance.

While GaAs materials offer numerous advantages, there are also some limitations associated with their use. One limitation is the higher cost of GaAs compared to other semiconductor materials, primarily due to the complex manufacturing processes involved. This cost factor restricts the widespread adoption of GaAs in certain applications where cost-effectiveness is a critical consideration. Additionally, GaAs is a brittle material, making it more prone to cracking and breakage during handling and fabrication. Moreover, GaAs-based devices may face challenges in scaling down to smaller dimensions due to material properties and technological constraints. Despite these limitations, ongoing research and advancements aim to address these issues and further enhance the capabilities and cost-effectiveness of GaAs materials for broader application domains.

\subsection{STEP 3: Optimization of Thickness and Carrier Density for GaAs layer}

Furthermore, figure 6 represents the proposed solar cell architecture which includes another layer of the p-GaAs layer to the baseline multi-junction solar cell. Subsequently, the GaAs layer is added on top of the p-CIGS layer next to the left contact (back) of the proposed solar cell architecture for analysing the electrical and electronic performance in the real-world environment with the help of the SCAPS-1D simulation tool. Moreover, the light in this scenario is incident from the right contact (front) to achieve the maximum possible efficiency of the proposed solar cell architecture. In addition to this, all working environments of the simulation setup were kept to same so as to avoid any discrepancy in the analysis of the output results and thus, the overall performance of the multi-junction solar cell.

Consecutively, the optimization of the GaAs semiconductor material was performed in terms of the thickness and doping concentration with the value ranging between 0.5 $\mu$m to 5.0 $\mu$m and $1\times10^{11}$ (1/cm) to $1\times10^{20}$ (1/cm), respectively. Accordingly, figure 7 showcases that the maximum efficiency and open circuit voltage of the proposed solar cell architecture is achieved at 45.47 and 1.16 V, respectively, at a thickness of 5.0 $\mu$m and acceptor density of GaAs material of $1\times10^{20}$ (1/cm). One of the most promising results in the heatmap confusion matrix indicates that the current density of the solar cell remains unaffected i.e., 43.88 ($A/m^2$) even by changing the thickness and doping concentration values of the solar cell. On the contrary, the percentage of the maximum fill factor is 89.52\% at the thickness and doping concentration values of 5.0 $\mu$m and $1\times10^{16}$ (1/cm), respectively.

Additionally, the current density (blue curve) as obtained from the IV curve outputs a peak value of 44.8 ($mA/Cm^2$) as shown in Figure 8 and thus, the maximum power observed is equivalent to the 46 W (the orange curve) at a thickness and doping concentration value of 5.0 $\mu$m and $1\times10^{20}$ (1/cm), respectively. Followed by, the maximum quantum efficiency achieved for the proposed solar cell with GaAs with the most optimized values of thickness and doping concentration is equal to 99.2 \% as shown in Figure 8. It is worth mentioning that the quantum efficiency curve consists of the quantum efficiency (\%) vs the wavelength curve (nm) with a range between 300 (nm) to 1130 (nm).

\section{Discussions}

In this section of the manuscript,  we discuss the results of the comparison of various IV, PV and QE characteristics of the proposed baseline solar cell, thickness optimization curve, and doping concentration curve and a thorough comparison is made for the electrical properties of the solar cell after adding the p-GaAs layer on top of the proposed baseline solar cell architecture. The comparison gives a clear indication of the most optimized thickness and doping concentration values that need to be set whilst manufacturing the different solar cell architectures at a mass level.

\begin{figure}[!t]
 \centering
 \includegraphics[width=8cm]{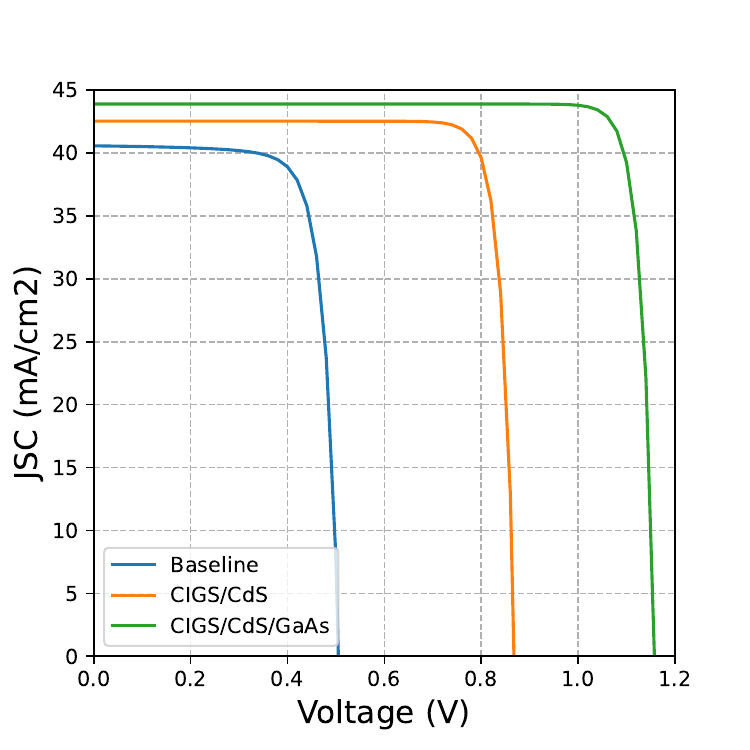}
 \caption{Figure represents a critical comparison of IV Characteristics of the proposed baseline multi-junction solar cell (p-CIGS/n-CdS/n-ZnO) architecture after the thickness optimization (blue). Followed by, the orange curve shows the results after the optimization of the doping concentration of the proposed solar cell. Lastly, the grey curve shows the results of the optimization of both thickness and doping concentration after introducing the GaAs layer to the proposed solar cell architecture.}
 \label{fig:Ref11}
\end{figure}

\subsection{Comparison of IV Characterisitcs}

Figure 9 represents the IV characteristics in terms of the current density, JSC ($mA/cm^2$) of the baseline - after thickness optimization (blue), the optimized curve for doping concentration (orange) and the optimization results after introducing the GaAs layer (green) on the proposed solar cell with varied values of the open circuit voltage, i.e., 0.5V, 0.83V and 1.17V, respectively. The curve indicates that the optimization of the solar cell architecture along with the thickness and current density led to an increase in the open circuit voltage of the solar cell architecture. In addition, an enhancement in a solar cell's performance or a change in its operating circumstances is often indicated by an increase in the VOC of the solar cell on the IV curve. This improvement of VOC is attributable to circumstances like better material quality, less charge carrier recombination, lower series resistance, or different operating environment like temperature. However, to appropriately evaluate the effectiveness and overall performance of the solar cell, it's crucial to take into account the entire IV curve, including elements like the short circuit current (Isc), fill factor (FF), and power output which is discussed in the subsequent subsection of the manuscript.

\begin{figure}[!t]
 \centering
 \includegraphics[width=8cm]{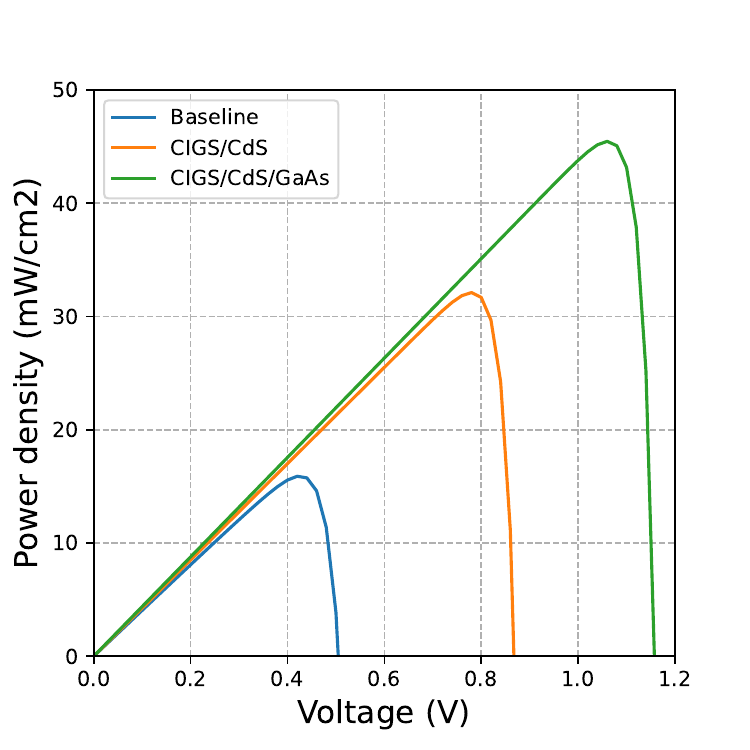}
 \caption{Figure represents the comparison of PV Characteristics of the proposed baseline multi-junction solar cell (p-CIGS/n-CdS/n-ZnO) architecture after the thickness optimization (blue). The orange curve shows the results after the optimization of the doping concentration of the proposed solar cell and the grey curve shows the results of the optimization of both thickness and doping concentration after introducing the GaAs layer to the solar cell.}
 \label{fig:Ref12}
\end{figure}

\subsection{Comparison of PV Characterisitcs}

Another comparison of power output is presented in Figure 10 showing a relation of the power density ($mW/cm^2$) vs voltage (V) characteristics, where like the previous step, the thickness optimization (blue), doping concentration (orange) and the p-GaAs layer optimization (green) are evaluated. The maximum of the power curve, represented as $P_{MP}$, is the point at which the solar cell should be operated to produce the most electricity. It occurs at a voltage of $V_{MP}$ and a current of $I_{MP}$ and is also referred to as $P_{MAX}$ or maximum power point (MPP). Figure 10 showcases that the power output for the thickness optimization of the proposed baseline solar cell architecture has a peak of 25 W, whereas that of the doping concentration is measured as 32 W. On the contrary, introducing the p-GaAs layer on the top of the proposed solar cell yields a peak value of power at 45.47 W. 

Therefore, adding a p-GaAs layer highly affects the output power of the solar cell and accordingly, it also results in an increased highest efficiency of the solar cell to 45.47\% which is thus so far the maximum efficiency recorded for a multi-junction solar cell consisting of p-GaAs/p-CIGS/n-CdS semiconductor materials. Moreover, the solar cell shows an overall improvement in functionality and effectiveness. The increased power output is due to the solar cell's material quality has improved, allowing for better light absorption and electrical conversion with optimized thickness and doping concentration values performed on the SCAPS-1D simulation tool. Additionally, improved operating conditions, such as temperature and illumination levels, decreased recombination losses and lower series resistance are the other miscellaneous reasons that contribute to an increase in the overall power output.

\begin{figure}[!t]
 \centering
 \includegraphics[width=8cm]{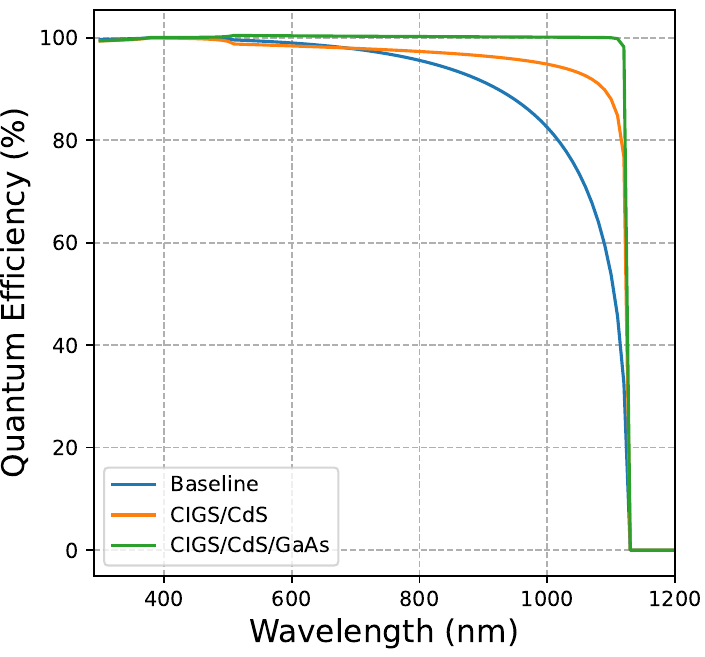}
 \caption{Figure showcases the results of the comparison of Quantum Efficiency (QE) for the proposed solar cell (p-CIGS/n-CdS/n-ZnO) architecture after the thickness optimization (blue). Accordingly, the orange curve shows the results after the optimization of the doping concentration of the proposed solar cell and the grey curve depicts the results of the optimization of both thickness and doping concentration after introducing the GaAs layer to the solar cell.}
 \label{fig:Ref13}
\end{figure}

\subsection{Comparison of QE Characterisitcs}

Subsequently, an improvement in the proposed solar cell's overall capacity to convert photons of various wavelengths into the electrical current is evaluated using the quantum efficiency vs. lambda (wavelength) curve. The solar cell is growing more effective at absorbing a larger variety of photons across the electromagnetic spectrum as QE rises. The increased light-trapping techniques, improved material characteristics, decreased recombination losses, and improved device topologies are one of the major contributing factors that led to a sudden rise in the overall quantum efficiency. Therefore, an increase in the quantum efficiency (\%) in the quantum efficiency vs Lambda (wavelength in nm) curve depicts that the solar cell is performing efficiently in regards to capturing a wide range of wavelength spectrum which further results in an exponential rise in the total power conversion efficiency of the proposed solar cell. Accordingly, the authors presented a comparative analysis of the quantum efficiency (\%) for baseline thickness optimization (blue), doping concentration optimization (orange) and the optimization after introducing the GaAs layer. As analysed from Figure 11, the quantum efficiency for baseline thickness, doping concentration and the GaAs layer is 82\%, 94\% and 99.48\% at the 1020 nm of wavelength.


\section{Conclusions}

This paper proposes an efficient three-layered p-GaAs/p-CIGS/n-CdS (PPN), a unique solar cell architecture. Copper indium gallium selenide (CIGS)-based solar cells exhibit substantial performance than the ones utilizing cadmium sulfide (CdS). On the contrary, CIGS-based devices are more efficient, considering their device performance, environmentally benign nature, and reduced cost. Therefore, our paper proposes a numerical analysis of the homojunction PPN-junction GaAs solar cell structure along with n-ZnO front contact that was simulated using the Solar Cells Capacitance Simulator (SCAPS-1D) software. Moreover, we investigated optimization techniques for evaluating the effect of the thickness and the carrier density on the performance of the PPN layer on solar cell architecture. Subsequently, the paper discusses the electronic characteristics of adding GaAs material on the top of the conventional (PN) junction, further leading to improved values of the parameters, such as the power conversion efficiency (PCE), open-circuit voltage (VOC), fill factor (FF) and short-circuit current density (JSC) of the solar cell. The most promising results of our study show that adding the GaAs layer using the optimised values of thickness as 5 ($\mu$m) and carrier density as $1\times10^{20}$ (1/cm) will result in the maximum PCE, VOC, FF, and JSC of 45.7\%, 1.16 V, 89.52\% and 43.88 $(mA/m^{2})$, respectively, for the proposed solar cell architecture.

This paper proposes an efficient three-layered p-GaAs/p-CIGS/n-CdS (PPN), a unique solar cell architecture. Copper indium gallium selenide (CIGS)-based solar cells exhibit substantial performance than the ones utilizing cadmium sulfide (CdS). On the contrary, CIGS-based devices are more efficient, considering their device performance, environmentally benign nature, and reduced cost. Therefore, our paper proposes a numerical analysis of the homojunction PPN-junction GaAs solar cell structure along with n-ZnO front contact that was simulated using the Solar Cells Capacitance Simulator (SCAPS-1D) software. Moreover, we investigated optimization techniques for evaluating the effect of the thickness and the carrier density on the performance of the PPN layer on solar cell architecture. Subsequently, the paper discusses the electronic characteristics of adding GaAs material on the top of the conventional (PN) junction, further leading to improved values of the parameters, such as the power conversion efficiency (PCE), open-circuit voltage (VOC), fill factor (FF) and short-circuit current density (JSC) of the solar cell. The most promising results of our study show that adding the GaAs layer using the optimised values of thickness as 5 ($\mu$m) and carrier density as $1\times10^{20}$ (1/cm) will result in the maximum PCE, VOC, FF, and JSC of 45.7\%, 1.16 V, 89.52\% and 43.88 $(mA/m^{2})$, respectively, for the proposed solar cell architecture.

\bibliographystyle{IEEEtran}
\bibliography{ref}

\end{document}